\documentclass[aps,prb,reprint,showpacs,twocolumn,longbibliography,superscriptaddress]{revtex4-2}

\usepackage[dvips]{graphicx}
\usepackage{dcolumn}% Align table columns on decimal point
\usepackage{bm}% bold mat/cite
\usepackage{xspace}
\usepackage{multirow}
\usepackage{natbib}
\usepackage{color}

\begin{document}

\preprint{}
\title{Possible Topological Superconductivity in a Topological Crystalline Insulator (Pb$_{1-x}$Sn$_x$)$_{1-y}$In$_y$Te}
\author{I. Pletikosi\'{c}}
\affiliation{Condensed Matter Physics and Materials Science Department, Brookhaven National Lab, Upton, New York 11973, USA}
\affiliation{Department of Physics, Princeton University, Princeton, NJ 08544, USA}
\author{T. Yilmaz}
\affiliation{National Synchrotron Light Source, Brookhaven National Lab, Upton, New York 11973, USA}
\author{B. Sinkovic}
\affiliation{Department of Physics, University of Connecticut, Storrs, Connecticut 06269, USA\\}
\author{A. P. Weber}
\affiliation{National Synchrotron Light Source, Brookhaven National Lab, Upton, New York 11973, USA}
\affiliation{ICFO-Institut de Ciencies Fotoniques: Barcelona, Spain}
\author{G. D. Gu}
\affiliation{Condensed Matter Physics and Materials Science Department, Brookhaven National Lab, Upton, New York 11973, USA}
\author{T. Valla}
\affiliation{Condensed Matter Physics and Materials Science Department, Brookhaven National Lab, Upton, New York 11973, USA}
\affiliation{Donostia International Physics Center, 20018 Donostia-San Sebastian, Spain}
\email{tonica.valla@dipc.org}

\date{\today}
\begin{abstract}
Superconductivity in topological insulators is expected to show very unconventional features such as $p+ip$ order parameter, Majorana fermions etc. However, the intrinsic superconductivity has been observed in  a very limited number of materials in which the pairing symmetry is still a matter of debate. Here, we study the topological crystalline insulator (Pb$_{1-x}$Sn$_x$)$_{1-y}$In$_y$Te, where a peculiar insulator to superconductor transition was previously reported near the gap inversion transition, when the system is nearly a 3-dimensional Dirac semimetal. Both the existence of superconductivity near the 3-dimensional Dirac semimetal and the occurrence of insulator to superconductor transition in an isotropic material is highly unusual. We suggest that the observed phenomena are related to an intrinsic instability of a 3-dimensional Dirac semimetal state  in (Pb$_{1-x}$Sn$_x$)$_{1-y}$In$_y$Te and "flattening" of the bulk valence and conduction bands as they acquire a Mexican hat-like dispersion on the inverted side of the phase diagram. This favors the pairing instability if the chemical potential is pinned to these flat regions.

\end{abstract}
\vspace{1.0cm}


\maketitle

A 3-dimensional (3D) Dirac point with linear dispersion in all three directions can be created at a time-reversal invariant momentum (TRIM) by tuning an external parameter $m$, if the valence and conduction bands have the opposite parities \cite{ShuichiMurakami2007,Murakami2007,Yang2014a}. This occurs, for example, at a quantum critical point ($m = m_c$) between a normal insulator and a Z2 topological insulator, or a topological crystalline insulator. In the gap-inverted (topological) phase, the topological states that unavoidably exist at boundaries \cite{VolkovB.A.1985,Fradkin1986}, have been in the focus of both theoretical and experimental studies over the last $\sim$ 2 decades \cite{Murakami2004,Kane2005,Bernevig2006,Konig2007,Fu2007a,Noh2008a,Konig2008a,Hsieh2008,Fu2011a,
Xu2012,Liu2013a,Yang2014,Pletikosic2014a,Gyenis2013a}.
The approach to the 3D Dirac quantum critical point requires a fine-tuning of an external 
parameter ($m$) such as pressure, or material composition that drives the transition \cite{Svane2010,Suchalkin2020a}. In general, the 3D Dirac point with four-fold degeneracy does not have a topological number and it is therefore unstable against small perturbations. However, it was later realized that particular lattice symmetries may result in the band structure with stable 3D Dirac cones and shortly after the theoretical prediction \cite{Wang2012a,Wang2013a} several materials including Na$_3$Bi and Cd$_3$As$_2$ were experimentally identified as 3D Dirac semi-metals (DSM) \cite{Liu2014a,Borisenko2014,Neupane2014}. 
While the 3D Dirac points in these 3D DSMs are protected by crystal 
symmetries, the generally unprotected 3D Dirac points at the 
band inversion transitions could potentially display a very rich physics. A particularly interesting example is In-doped Pb$_{1-x}$Sn$_x$Te a topological crystalline insulator (TCI), that shows an intriguing transition from a relatively good insulator to a robust 
superconductor as a function of In concentration when in the vicinity of a transition between a trivial 
and topological crystalline insulator at $x=x_c\approx 0.36$ \cite{Bushmarina1991,Kozub2006,Zhong2014,Zhong2015,Zhong2017,Sapkota2020}. The insulator to superconductor transitions have been usually observed in thin films and in some quasi-2D materials as a consequence  of strong quantum phase fluctuations due to reduced dimensionality \cite{Lin2015,Chudnovskiy2022}. In a relatively isotropic 3D material, such transition is highly 
unexpected. The appearance of superconductivity itself is also highly unusual in a system where the low-energy quasiparticles resemble 3D Dirac fermions with linearly vanishing density of states. 

Another interesting aspect of In-doped Pb$_{1-x}$Sn$_x$Te is the posibility that its superconducting phase represents an intrinsic topological superconductor \cite{Fu2008}. While for a long time the focus has been on Sr$_2$RuO$_4$, considered to be an intrinsic chiral $p$-wave triplet superconductor \cite{Rice1995,Ishida1998,DasSarma2006}, more-recent  results excluded that possibility \cite{Pustogow2019}. Currently, the number of bulk topological superconductors is very limited, with Cu$_x$Bi$_2$Se$_3$ and Sn$_{(1-x)}$In$_x$Te being the rare candidates, with $T_c$ around $2-4$ K \cite{Hor2010,Erickson2009a,Sasaki2012}.  Cu$_x$Bi$_2$Se$_3$ is found to be very inhomogeneous, with a small superconducting volume and therefore is not ideal for spectroscopic studies that could determine the order parameter or the existence of Majorana fermions. In more-homogeneous Sn$_{(1-x)}$In$_x$Te, spectroscopic studies have provided evidence for the odd-parity pairing and topological superconductivity \cite{Sasaki2012}. More recently, several candidates for topological superconductors are discovered in materials that make natural superlattices involving rock-salt Pb$_{1-x}$Sn$_x$Te layers, in which the specific heat suggests the existence of gap nodes \cite{Sasaki2014,Luo2016a}. 

%%%%fig 1 
%#######################################################################
\begin{figure*}[htbp]
\begin{center}
\includegraphics[width=14cm]{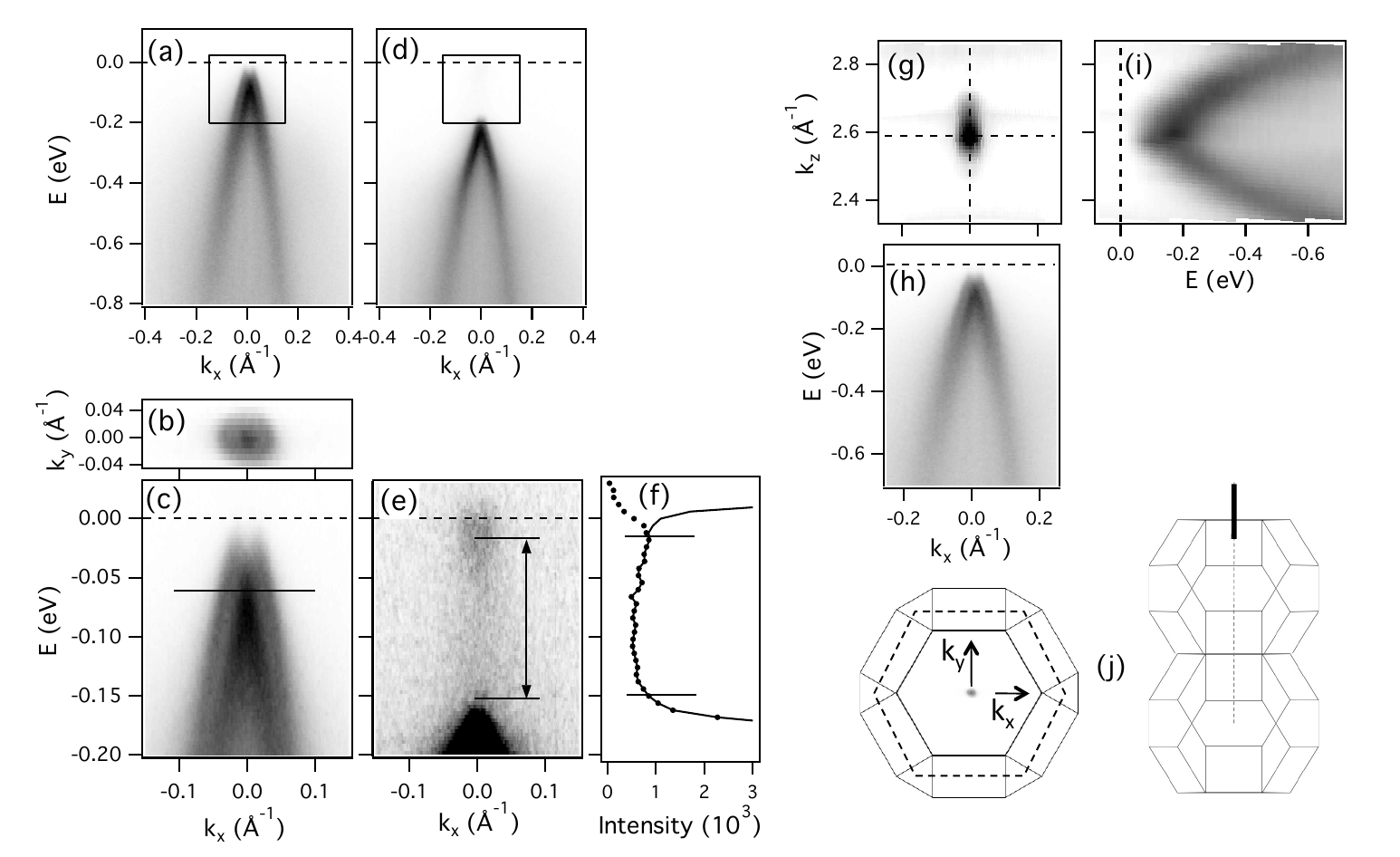}
\caption{Electronic structure of a thick (111) Pb$_{0.85}$Sn$_{0.15}$Te film. (a) 
Valence band dispersion along the $\bar\Gamma-\bar K$ momentum line of the surface Brillouin zone (SBZ) for the Te-
terminated surface. (b) Constant energy contour of the ARPES intensity at $E=-60$ meV (marked in (c) with the line) as a function of the in-plane momentum. (c) Close-up of the spectral region marked by the rectangle in (a). (d) and (e) show the corresponding spectra for the Pb-Sn terminated surface. (f) Energy distribution curve (EDC) at $k_x=0$ from (e). The raw intensity and that divided by the Fermi distribution is plotted as dot-markers and solid line, respectively. The spectra in (a-f) were recorded at $\hbar\nu=17$ eV, corresponding to $k_z=2.6$ \AA ($L$ point in between the 2$^{nd}$ and 3$^{rd}$ bulk BZ). The arrow in (e) represents the gap between the VB and CB, $\Delta_L\approx$ 130 meV. (g) 
ARPES intensity for the Te-terminated surface at $E=-60$ meV as a function of the $k_z$ along the $\Gamma-L-\Gamma$ of 
the BBZ and $k_x$ along the $\bar\Gamma\bar K$ line of the SBZ. (h) In-plane dispersion at 
$k_z=2.6$ \AA (same as (a)). (i) $k_z$ dispersion of the valence band along the $\Gamma-
L-\Gamma$ line of the bulk BZ. 
Intensity maps in (g,i) were produced from spectra taken at photon energies ranging from $\hbar\nu=12.5$ eV to $\hbar\nu=24$ eV. All the spectra were taken at $\sim50$ K. (j) Projection of the bulk BZ to the (111) surface and the SBZ 
(dashed hexagon) (left). The small circle at the $\bar\Gamma$ is the same contour from (b). The 1$^{st}$ and 
2$^{nd}$ bulk BZ with the probed region of $k_z$ marked by a thick line. 
}
\label{Fig1}
\end{center}
\end{figure*}
%#######################################################################

%fig 2ARPES %#######################################################################
\begin{figure}[htpb]
\begin{center}
\includegraphics[width=8cm]{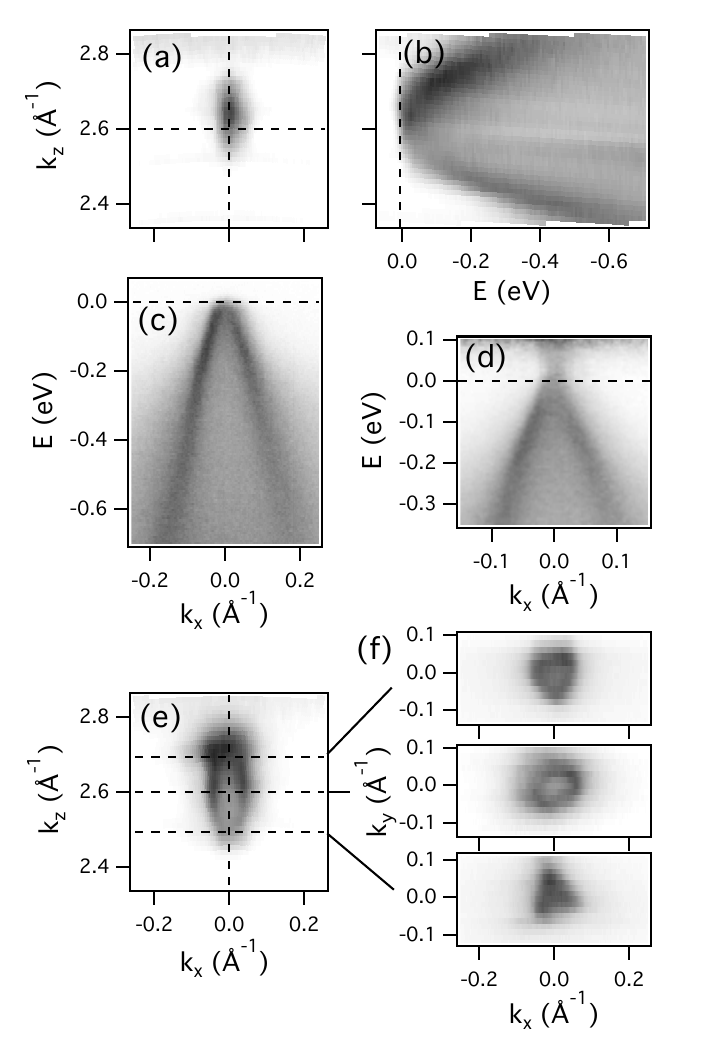}
\caption{3D Electronic structure of a thick (111) Pb$_{0.6}$Sn$_{0.4}$Te film. (a) 
ARPES intensity at the Fermi level (Fermi surface) as a function of the 
$k_z$ along the $\Gamma-L-\Gamma$ of the BBZ and $k_x$ along t
he $\bar\Gamma\bar K$ line of the SBZ.  
(b) $k_z$ dispersion of the valence band along the $\Gamma-L-\Gamma$ line, near the $L$ 
point between the $2^{nd}$ and $3^{rd}$ BBZ.
(c) Dispersion of the valence band along the $\bar\Gamma\bar K$ line of the SBZ, 
taken at 17 eV photon energy (corresponding to the $W-L-W$ line in the bulk BZ). (d) The same as in (c) but recorded at $\sim 300$ K and divided by the corresponding Fermi distribution. (e) Constant energy contour at $E=-100$ meV as a function of $k_x$ and $k_z$ ($k_y=0$). (f) Constant energy contours of in-plane ARPES intensity at $E=-100$ meV, taken at $\hbar\nu=15$ eV (bottom), $\hbar\nu=17$ eV, (middle) and $\hbar\nu=19$ eV (top), corresponding to three different $k_z$, as indicated by dashed lines in (e). All the spectra were taken at 80 K, except for the spectrum in (d). 
}
\label{Fig2}
\end{center}
\end{figure}
%#######################################################################

An alternative avenue towards topological superconductivity in heterostructures involving superconductors and topological insulators has been also explored \cite{Fu2008,Wang2012b,Yi2022}, including the TI films on cuprate high-temperature superconductors (HTSC). However, the studies claiming a large superconducting gap on the topological surface state on Bi$_2$Se$_3$ grown on Bi2212 \cite{Zareapour2012,Wang2013} were later shown to be false \cite{Yilmaz2014}. The complete absence of proximity effect seems to be intrinsic to HTSC - the films grown on cuprate superconductors donate electrons, thus making the interface so underdoped that superconductivity completely vanishes \cite{Kundu2020,Kundu2021}, making  HTSCs inadequate for inducing topological superconductivity. 

Here, we show that the topological crystalline insulator Pb$_{1-x}$Sn$_{x}$Te might be a good candidate for an intrinsic topological superconductor when doped with Indium. When near the gap inversion point, In-doped Pb$_{1-x}$Sn$_{x}$Te shows a transition between a real bulk TCI and a bulk superconductor \cite{Zhong2017}. By studying the details of electronic structure in angle-resolved photoemission spectroscopy (ARPES), we suggest that this peculiar transition is caused by the transformation of bulk bands from the linear, 3D-Dirac-like to "Mexican hat"-like ones on the topological side, where an inevitable flattening at the onset of that transformation must occur. This transformation could support superconductivity when the flat regions are brought near the chemical potential \textit{via} In-doping.

The ARPES experiments were carried out on a Scienta SES-R4000 electron spectrometer at 
beamline U5UA at the National Synchrotron Light Source (NSLS), in the photon energy range from 12.5 
to 62 eV and at the OASIS-laboratory at Brookhaven National Laboratory, using monochromatized He I$_{\alpha}$ radiation \cite{Kim2022}. The core level spectra were taken at NSLS using 110-150 eV photons. The total instrumental energy resolution was $\sim$ 6 meV at 12-24 
eV, $\sim$ 12 meV at 50-62 eV and $\sim$ 30 meV at 110-150 eV photon energy.  Angular resolution was better than $\sim 
0.15^{\circ}$ and $0.4^{\circ}$ along and perpendicular to the slit of the analyzer, respectively. 

%fig 3ARPES %#######################################################################
\begin{figure*}[htpb]
\begin{center}
\includegraphics[width=14cm]{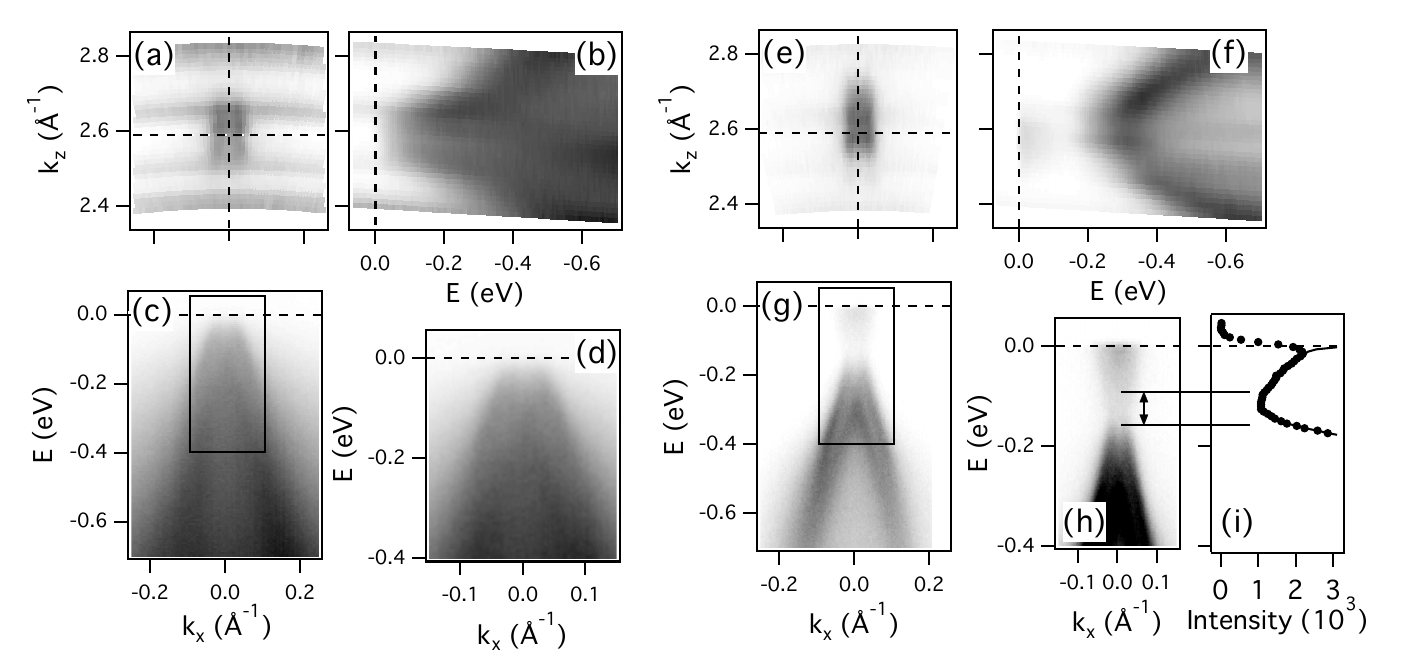}
\caption{3D Electronic structure of  Pb$_{0.7}$Sn$_{0.3}$Te and (Pb$_{0.7}$Sn$_{0.3}$)$_{0.84}$In$_{0.16}$Te. (a) 
ARPES intensity at the Fermi level (Fermi surface) as a function of the 
$k_z$ along the $\Gamma-L-\Gamma$ of the BBZ and $k_x$ along the $\bar\Gamma\bar K$ line of the SBZ for Pb$_{0.7}$Sn$_{0.3}$Te.  
(b) $k_z$ dispersion of the valence band along the $\Gamma-L-\Gamma$ line, near the $L$ 
point between the $2^{nd}$ and $3^{rd}$ BBZ.
(c) Dispersion of the valence band along the $\bar\Gamma\bar K$ line of the SBZ, 
taken at 17 eV photon energy (corresponding to the $W-L-W$ line of the bulk BZ). (d) Close-up of the spectral region marked with the rectangle in (c). (e-h) The corresponding panels for In-doped sample, (Pb$_{0.7}$Sn$_{0.3}$)$_{0.84}$In$_{0.16}$Te. The arrow in (h) represents the gap between the VB and CB, $\Delta_L\approx$ 65 meV.  (i) EDC at $k_x=0$ from (h). The raw intensity and that divided by the Fermi distribution is plotted as dot-markers and solid line, respectively. The spectra in (a-d) and (e-i) were taken at 80 and 50 K, respectively. 
}
\label{Fig3}
\end{center}
\end{figure*}
%#######################################################################

Thick, bulk-like (2-5 $\mu$m) films of (111) (Pb$_{1-x}$Sn$_x$)$_{1-y}$In$_y$Te were grown either on BaF$_2$ or Bi$_2$Te$_3$ substrates, using the open hot wall epitaxy method \cite{Strauss1967} and a single-source evaporators loaded with crushed single crystals of (Pb$_{1-x}$Sn$_x$)$_{1-y}$In$_y$Te \cite{Zhong2014}.
The BaF$_2$ substrates were cleaved in air and annealed in ultra high vacuum (UHV) at 600$^{\circ}$C for one hour, while the Bi$_2$Te$_3$ substrates were cleaved in UHV before the film deposition. During the growth, the substrate was kept at $\sim300^{\circ}$C. The thickness was determined by the quartz crystal thickness monitor. Composition of the films was checked by measuring photoemission from the shallow core levels (In $4d$, Sn $4d$, Te $4d$ and Pb $5d$) and comparing them with spectra of the source material that was broken \textit{in situ}. After ARPES experiments, films were checked for superconductivity by SQUID magnetometry. In several cases, the $T_c$ in the film was slightly higher than in the source material, with the maximum $T_c$ of $\sim$6 K. 

\subsection{Pure Pb$_{1-x}$Sn$_{x}$Te}
Fig. \ref{Fig1} shows the ARPES data from the film on the trivial side of the phase diagram: Pb$_{0.85}$Sn$_{0.15}$Te. The (111) surfaces can be either Te- or Pb-Sn-terminated and in the Pb-rich films ($x<0.3$), both terminations can be distinguished in ARPES, as shown in Fig. \ref{Fig1}. The panels (a-c) represent the in-plane electronic structure of the mostly Te-terminated surface, while the panels (d-e) represent mostly Pb-Sn termination. The two terminations could be distinguished as they display slightly different ratio between the Te $4d$ and Pb $5d$ (Sn $4d$) core levels. Both terminations show almost identical (aside from the $\sim 0.2$ eV energy shift) and fairly isotropic in-plane dispersion. Both also display bands split in a Rashba-like manner, near their extrema. The splitting is more obvious on the Te-terminated surface, indicating a stronger inversion symmetry breaking surface field that would result in larger spin-orbit interaction. However, the bands disperse rather strongly with $k_z$ (panels (g,i) and it is not clear if the Rashba-like split features are the surface states. Indeed, a suggestion that the bulk bands could also display the Rashba splitting when the breaking of the inversion symmetry due to the surface goes deep enough into the crystal was recently discussed in relation to a new candidate for spin-triplet superconductor, UTe$_2$ \cite{Ran2019,Yu2022}. Here, we note that the momentum width of the valence band along the $k_z$ ($\Delta k_{\perp}\sim0.06$\AA$^{-1}$) is comparable to the in-plane width, implying the surprisingly large ARPES probing depth at these photon energies.
The more-conventional possibility is that the observed Rashba-like split features are trivial precursors of the topological surface states. Their apparent dispersion with $k_z$, or alternatively, their existence only inside an extremely limited $k_z$ interval corresponding to the very vicinity of the $L$ point of the BZ, would suggest their strong mixing with the bulk states. Similar effects were observed on the topological side of Pb$_{1-x}$Sn$_{x}$Se \cite{Polley2014}. 

On the surface terminated mostly by Pb-Sn, the bottom of the bulk conduction band is visible at $\hbar\nu=17$ eV, enabling the determination of the direct band-gap $\Delta_L\approx 130$ meV at  50 K, in agreement with the early optical studies at similar compositions \cite{Dimmock1966,Takaoka1989}. The photon-energy dependence that gives the access to the out-of-plane electronic structure is shown in panels (g,i) for the mostly Te-terminated surface. The set of $k_y=0$ spectra measured at photon energies ranging from 12.5 to 23 eV is converted to $k_z$ using the free electron approximation for the final photoemission state, with $V_0=13$ eV for the \lq\lq{}inner\rq\rq{} potential. The emission at $\hbar\nu=17$ eV corresponds to $k_z=2.6$ \AA, the $L$ point between the second and the third bulk Brillouin zone (BBZ). The constant energy contours, shown for the Te-terminated surface (in-plane: panel (b), out-of-plane: panel (g)) correspond to $E=-60$ meV as there is no intensity at the Fermi level. The contour is elongated in the $k_z$ direction, in agreement with the band structure calculations \cite{Svane2010}. It also shows a limited span of the Rashba-like features in this direction. The in-plane dispersion (panel (h)) is approximately 2-3 times faster than the out-of plane one (panel (i)), and aside from a $\sim130$ meV gap, the electronic structure seems already very close to the 3D DSM. 

Fig.\ref{Fig2} shows the 3D electronic structure of a film on the TCI side, but very near the 
inversion transition: Pb$_{0.6}$Sn$_{0.4}$Te. At this composition, we no longer observe either Rashba-like split states or two distinct terminations. The reason is probably in the fact that with increasing $x$, the samples become significantly more hole-doped and for $x\geq0.4$, we only observe hole pockets with a very consistent sizes. The higher carrier concentration provides more effective screening of the surface potential, reducing the surface band bending. Consequently, the differences for different terminations should diminish. 
The Fermi surface (panel (a)) is a small ellipsoidal hole pocket, with $\sim 3:1$ ratio in $k_z$ over the in-plane size. If the pocket is the bulk feature, as the strong $k_z$ dispersion (panel (b)) indicates, and by assuming the ellipsoidal shape, we can estimate the bulk carrier concentration to be $n_h\sim 1.2\times10^{18}$ cm$^{-3}$ (four $L$ pockets in the BBZ containing $3\times10^{17}$ cm$^{-3}$ holes each). The Fermi velocities, deduced from momentum distribution curves (MDC) fitting are 4.4 eV\AA\ and 1.7 eV\AA\ for in- and out-of-plane directions, respectively. The TSS probably overlaps with these (bulk) states as was shown in $n$-type surfaces of Pb$_{1-x}$Sn$_{x}$Se \cite{Turowski2023,Pletikosic2014a,Gyenis2013a}. 

At higher energies, the contours become distorted and strongly trigonally warped, both below and above the $L$ point (panel (f)), with the direction of warping being opposite across the zone boundary. This was predicted in the band structure calculations, but never observed experimentally \cite{Svane2010}.
It seems from the spectra in Fig. \ref{Fig2} that this composition is indeed very close to the critical one at which the band inversion occurs. The bands are nearly linear and the Fermi surface is very small. Unfortunately, as the films at this composition are always $p$-type, we do not have  an access to the conduction band at low temperature. However, at room temperature, we could access it, and the spectrum (panel (d)) indicates that the gap is indeed negligible. 

\subsection{Effects of In Doping}

%fig 4ARPES 
%#######################################################################
\begin{figure*}[htbp]
\begin{center}
\includegraphics[width=14cm]{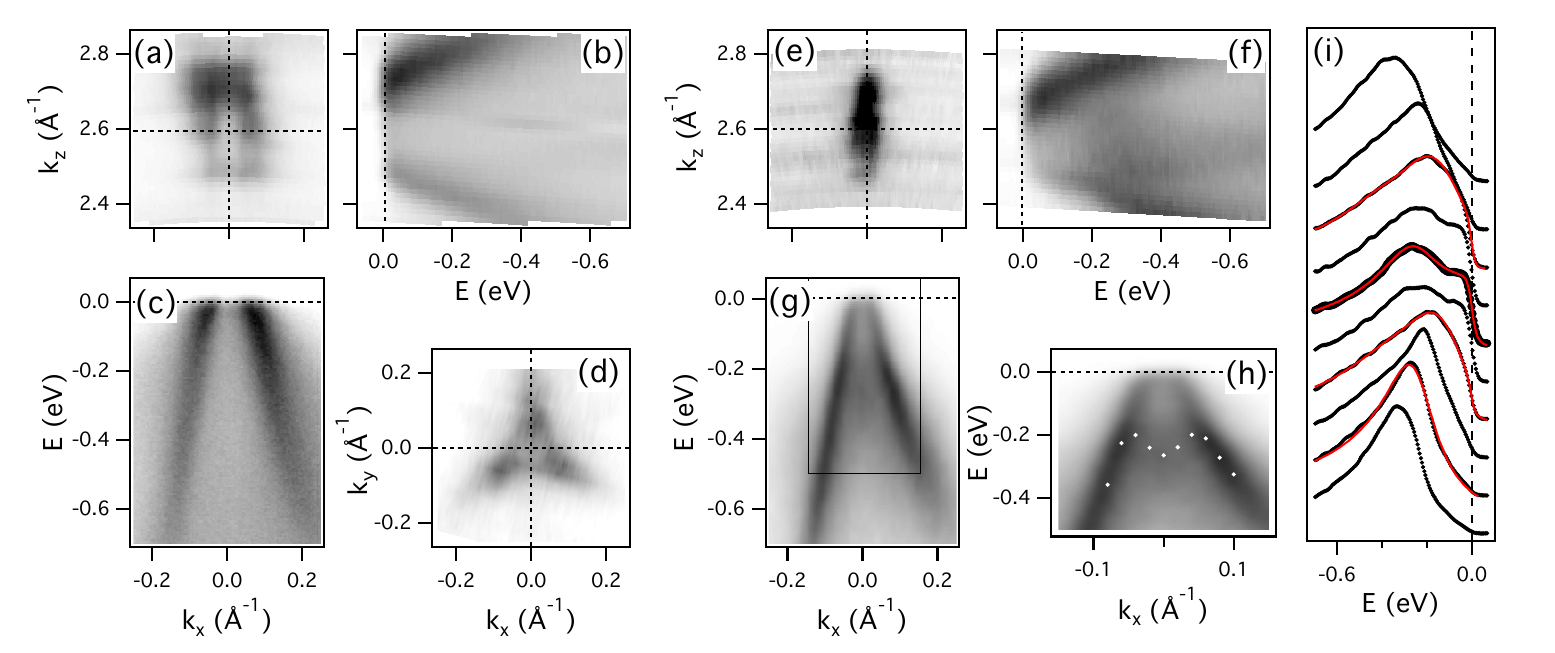}
\caption{3D Electronic structure of  Pb$_{0.5}$Sn$_{0.5}$Te and (Pb$_{0.5}$Sn$_{0.5}$)$_{0.7}$In$_{0.3}$Te.
(a) 
ARPES intensity at the Fermi level (Fermi surface) as a function of the 
$k_z$ along the $\Gamma-L-\Gamma$ of the BBZ and $k_x$ along the $\bar\Gamma\bar K$ line of the SBZ for Pb$_{0.5}$Sn$_{0.5}$Te.  
(b) $k_z$ dispersion of the valence band along the $\Gamma-L-\Gamma$ line, near the $L$ 
point between the $2^{nd}$ and $3^{rd}$ BBZ.
(c) Dispersion of the valence band along the $\bar\Gamma-\bar K$ line of the SBZ, 
taken at 17 eV photon energy (corresponding to the $W-L-W$ line of the  BBZ). (d) Constant 
energy contour of in-plane ARPES intensity at $E=0$ (Fermi surface) taken at $\hbar\nu=17$ eV, corresponding to $k_z=2.6$ \AA. (e-g) The same as (a-c), but for In-doped sample, (Pb$_{0.5}$Sn$_{0.5}$)$_{0.7}$In$_{0.3}$Te. All the spectra were taken at 80 K. (h) Close-up of the spectral region marked with the rectangle in (g). The white points indicate dispersion (peak position) of the bulk valence band from fitting the EDCs in (i). (i) EDCs of the spectral intensities from (g, h) (black) with selected fitting curves (red). The thick spectrum corresponds to the $k_x=0$, or the $L$ point in the bulk BZ. 
}
\label{Fig4}
\end{center}
\end{figure*}
%#######################################################################

Fig. \ref{Fig3} shows the ARPES results for the trivial side of the (Pb$_{0.7}$Sn$_{0.3}$)$_{1-y}$In$_{y}$Te phase diagram, $x = 0.3$, with two different In concentrations, $y=0$ and $y=0.16$. The valence band in Fig. 3(c, d) again displays two states split in $k$ in the Rashba-like manner. If, as previously discussed, these are the surface states peaking in intensity at the $L$ point of the bulk BZ, they would indicate that the material is still on the trivial side of the phase diagram, and although they might have a significant contribution in transport of thin samples  \cite{Zhong2015}, they are not protected by any topological invariance. In that sense, they are very similar to the conventional Rashba surface states on Au(111). Increasing the In concentration to $y=0.16$ in Fig. 3(b) clearly shifts the Fermi level towards the bulk conduction band and now the Rashba-like surface states can be seen on both the conduction side and the valence side, forming a double-cone like shape, with a small gap still present between the two sides. This shows that at this composition, the material is very close to the critical point, at which the two cones should collapse into a single, un-gapped one—and the topological surface states should be fully formed. A previous ARPES study has provided evidence for topological surface states at $x = 0.4$ (without In) \cite{Xu2012}. The crucial point here is that the (Pb$_{0.7}$Sn$_{0.3}$)$_{1-y}$In$_{y}$Te does not become superconducting for any amount of In-doping \cite{Bushmarina1991,Zhong2015,Zhong2017} up to the In solubility limit. On the contrary, initially metallic, Pb$_{0.7}$Sn$_{0.3}$Te becomes a bulk insulator with In-doping. However, at low temperature, the resistivity becomes dominated by the contribution from the surface states and saturates in thin samples \cite{Zhong2015,Zhong2017}, a clear sign that even the trivial surface states can significantly participate in transport in certain cases. 

Fig. \ref{Fig4} shows the effects of In doping on the electronic properties of Pb$_{0.5}$Sn$_{0.5}$Te, on the gap inverted, TCI side of the phase diagram. Without In, ARPES indicates a heavily hole-doped system with the carrier concentration $n_h\sim 5\times10^{19}$ cm$^{-3}$ (4 times $1.25\times10^{19}$ cm$^{-3}$), almost two orders of magnitude larger than in the Pb$_{0.6}$Sn$_{0.4}$Te sample. In-plane Fermi velocities show significant anisotropy, ranging from $\sim 2.2$ eV\AA\ to $\sim 6$ eV\AA.  The slower states are likely bulk, while the fast dispersing ones are the TSS, as their velocity is similar to those spanning the gap in the In-doped case from Fig. \ref{Fig4}(g.h). The out of plane Fermi velocity is $\sim 2$ eV\AA.
When doped with In, the  $n_h$ dramatically decreases and, for (Pb$_{0.5}$Sn$_{0.5}$)$_{0.7}$In$_{0.3}$Te, the chemical potential is shifted by $\sim200$ meV, from inside of the valence band to near the bottom of conduction band (note that at this composition without indium, the inverted gap would be around 100 meV \cite{Heremans2012}). 
This shift in chemical potential seems to be in a nice agreement with transport in (Pb$_{0.5}$Sn$_{0.5}$)$_{1-y}$In$_{y}$Te which displays rather interesting behavior: initially metallic ($y=0$), the material becomes less conductive and shows an insulating bulk conductance at low In concentration ($y\approx0.06$). However, at even higher In concentrations ($y\ge0.1$), the material becomes metallic again, this time with the superconducting ground state \cite{Zhong2014,Zhong2017}. This non-monotonic behavior, together with our spectroscopic results, would suggests that the carriers, initially holes, turn into electrons with sufficient In doping. This is similar to what has been observed in Bi-doped material where the measured Hall coefficients support the transition from the hole to the electron conduction, albeit at much lower nominal dopant concentrations \cite{Volobuev2017}.

Furthermore, the bulk valence band, now fully occupied, displays a local minimum at $L$ point and two nearby maxima  as it disperses along the $k_x$ (Fig. \ref{Fig4}(h, i)). Its dispersion is obtained from the peak position of the higher binding energy peaks in EDCs, as represented by white points in panel (h).
%Dispersion along the $\Gamma-L$ line ($k_z$), seems to keep a simple shape, with a single maximum at $L$ point ((Fig. \ref{Fig4}(f)).

\begin{figure}[htbp]
\begin{center}
\includegraphics[width=8cm]{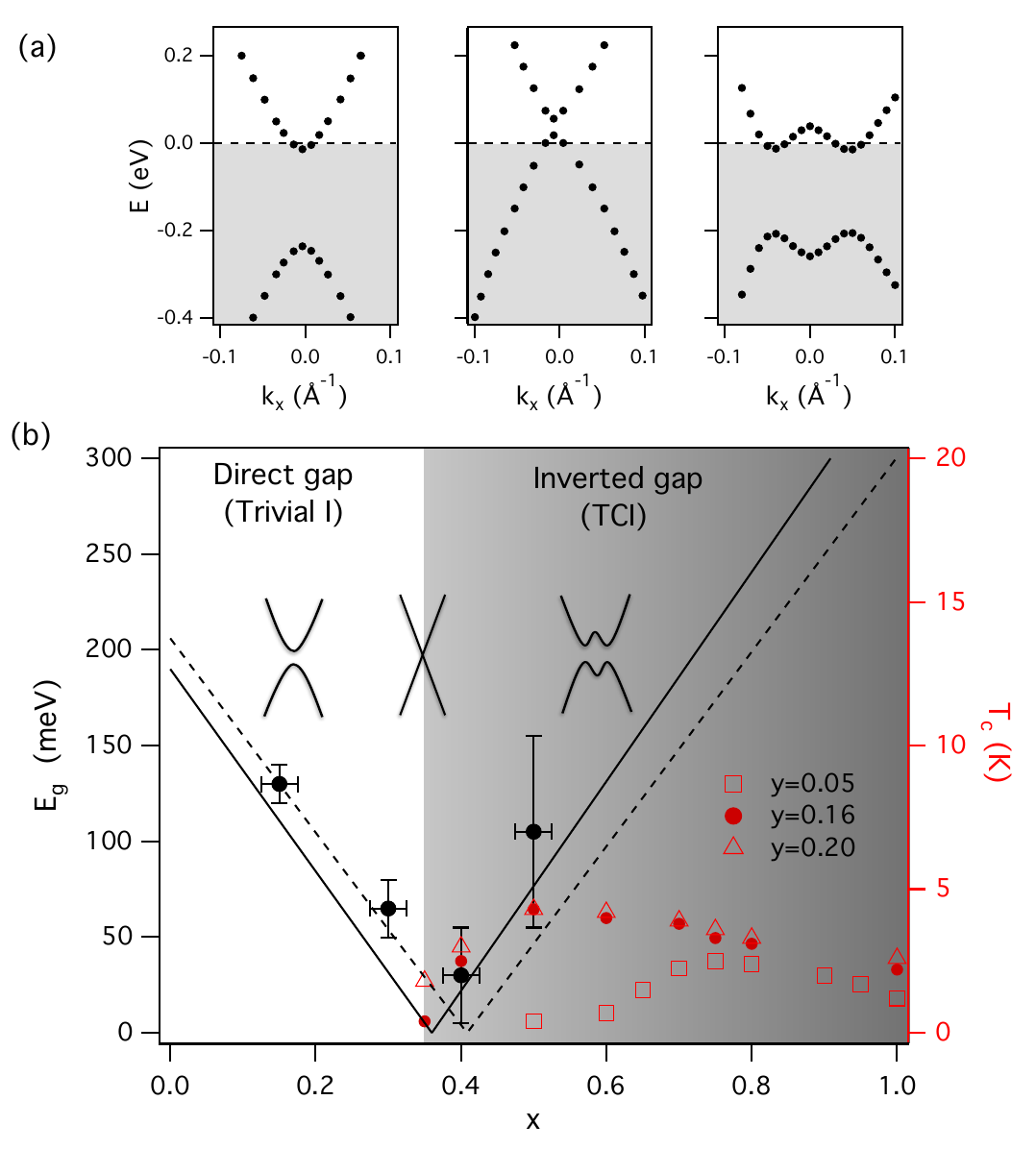}
\caption{Phase diagram of (Pb$_{1-x}$Sn$_x$)$_{1-y}$In$_y$Te. (a) Experimentally obtained in-plane dispersions of occupied (tinted) bulk states in Pb$_{0.85}$Sn$_{0.15}$Te (left), Pb$_{0.6}$Sn$_{0.4}$Te (middle) and (Pb$_{0.5}$Sn$_{0.5}$)$_{0.7}$In$_{0.3}$Te (right). The fitting of momentum distribution curves (MDCs) was performed for the first two cases, while EDC fitting was used in the third case. The unoccupied dispersions are inverted copies of the occupied ones. (b) The phase diagram of  (Pb$_{1-x}$Sn$_x$)$_{1-y}$In$_y$Te. The solid (dashed) line represents the fundamental gap, $E_g$, at $T=0$ ($T=50$ K) in Pb$_{1-x}$Sn$_x$Te from Heremans \textit{et al}  \cite{Heremans2012}. The black circles are gap estimates from our measurements, as discussed in the main text. Superconducting $T_c$ data points for In-doped Pb$_{1-x}$Sn$_x$Te (red symbols) are taken from ref. \cite{Bushmarina1991}. 
}
\label{Fig5}
\end{center}
\end{figure}

So, the question is why is the In-doped Pb$_{1-x}$Sn$_x$Te superconducting only on the gap-inverted, TCI side of the phase diagram \cite{Bushmarina1991}? After all, pure Pb has a higher $T_c$ than Sn and one could naively expect that the superconductivity will be stronger near the $x=0$. We think that the answer lies in the different shape of the bands before and after the gap inversion, as presented in Fig. \ref{Fig5}. On the trivial side, the valence and conduction bands have a single extremum along all three $k$-axis (Fig\ref{Fig1}). On approaching the critical point at $x\approx0.36$, these extrema approach each other and get lighter until they touch in a single (3D Dirac) point. On the topological side, and far from the inversion point, the bands have so called a Mexican hat band (MHB) shape \cite{Tung1969,He2014,Yang2014,Xu2020}. Specifically, in the case of SnTe, both the valence and conducting bands have a single maximum (minimum) at L point as they disperse along the $\Gamma-L$ line, whereas in the perpendicular direction, or within the (111) plane, the bands are  MHB, as realized early on by Tung \textit{et al}  and Rabii \cite{Rabii1969,Tung1969}. This is consistent with our experimental results (Fig. \ref{Fig4} (e-j)). The density of states (DOS) near the tops (bottoms) of such MHB can be dramatically enhanced relative to the band extrema on the trivial side, with the details depending on the models and dimensionality of the system \cite{He2014,Yang2014,Xu2020}. Moreover, the transition from a 3D Dirac (linear) bands to the MHB involves not only mass generation (gap opening), but also a flattening related with a transition from a single extremum to a MHB, that can further enhance the DOS and lead to singularities even in 3D systems \cite{He2014}. We argue that this will necessarily have an augmenting effect on the properties of a material dependent on DOS, as it will enhance electronic correlations, especially if the enhanced DOS can be brought near the chemical potential. In particular, we can expect that the electron-phonon coupling would be enhanced, eventually leading to superconductivity. The flat bands near the Fermi level are generally unstable against other symmetry-breaking phases, such as charge density waves and magnetism, but in In-doped Pb$_{1-x}$Sn$_x$Te, the superconductivity is the ground state tied to the MHB. Moreover, since the MHBs are inverted, the materials that possess them are excellent candidates for intrinsic topological superconductors. 

%fig 5
%#######################################################################
%#######################################################################

We note that both the gap magnitudes and the details in dispersions of valence/conduction bands shown in Fig. \ref{Fig5} depend on the analysis. They also depend on temperature, both in the theoretical and experimental studies. Our clearest cases are Pb$_{0.85}$Sn$_{0.15}$Te and (Pb$_{0.7}$Sn$_{0.3}$)$_{0.84}$In$_{0.16}$Te, where both valence and conduction bands are visible, suggest that the most appropriate method for gap determination is by the steep onsets of emission at band extrema  (Fig. \ref{Fig1}(e,f) and Fig. \ref{Fig3}(h,i)). 
For the remaining samples, the gap determination is less reliable. For example, in Pb$_{0.6}$Sn$_{0.4}$Te, the room temperature measurements displayed no detectable gap and we set the value of $\sim 30$ meV as the upper limit for low-temperature gap. 
In the pure Pb$_{0.3}$Sn$_{0.7}$Te sample, the conduction band is clearly out of the ARPES reach (Fig. \ref{Fig3}(c,d)). However, we were able to detect its threshold at room temperature after depositing potassium on the surface. The extracted gap was $\sim$150 meV, much larger than the one seen in In doped sample at low temperature, but when adjusted for temperature, by using the known thermal coefficients, the gaps were almost identical ($\sim65$ meV at 50 K). 
Also, the room temperature spectrum of the Pb0.4Sn0.6Te sample shows essentially no gap, but we have set the 30 meV as an upper limit at the low temperature. 
In the (Pb$_{0.5}$Sn$_{0.5}$)$_{0.7}$In$_{0.3}$Te case, the gap is determined from the dispersion of the peak maximum of the valence band and the assumption that the chemical potential is near the bottom of the conduction band (Fig.\ref{Fig4}(h,i)). From the Pb$_{0.85}$Sn$_{0.15}$Te case, we know that the distance between the sharp onset of emission and the peak maximum can be quite significant ($\sim 50$ meV) which would lead to a large overestimate for the gaps determined from peak positions. Therefore, in Fig. \ref{Fig5}, we have adjusted the gap magnitudes obtained from peak dispersions to those that would correspond to the onsets of emission, both for $x=0.15$ and $x=0.5$. Also, there is some ambiguity in how the gap is actually defined in the inverted case, where the bands display the MHB dispersion. However, the most important finding of our study, that the In-doped Pb$_{1-x}$Sn$_x$Te displays the MHB dispersion on the TCI side of the phase diagram, where superconductivity occurs, does not depend on those details.

On the other hand, the question whether the superconductivity in (Pb$_{1-x}$Sn$_{x}$)$_{1-y}$In$_{y}$Te is conventional or topological is still open. Topological superconductors are accompanied by gapless states at their boundaries, and they may be composed of Majorana fermions. The point-contact spectroscopy on In-doped SnTe suggests the odd-parity pairing and topological superconductivity \cite{Sasaki2012}, but the specific heat reveals a fully gapped superconductivity that would argue against it \cite{Novak2013}.
Similarly, the STM measurements of the superconducting gap in (Pb$_{0.5}$Sn$_{0.5}$)$_{0.7}$In$_{0.3}$Te on the (001) surface suggest that the sample seems to be fully gapped, without in-gap states—contrary to what would be expected for a topological superconductor \cite{Du2015}. However, we note that those measurements were performed on (100) surfaces of bulk crystals that poorly cleave. In addition, topological surface states on the (100) surfaces are different from those on the (111) surfaces studied here and their ability to support the zero modes might also be different \cite{Liu2013a}. Therefore, it would be highly desirable to perform STM studies on the (111) surfaces. Our ARPES results unambiguously show that the superconductivity in In-doped Pb$_{1-x}$Sn$_x$Te occurs not only on the topological side of the phase diagram, but also because of it, with the inverted MHBs playing the decisive role. Therefore, the In-doped Pb$_{1-x}$Sn$_x$Te is a TCl (for $x\ge 0.36$) and is also a superconductor, but it remains to be seen whether it is a topological superconductor with an odd order parameter and in-gap states. We hope that our results will stimulate further studies capable of answering these questions directly.

\begin{acknowledgments}
This work was supported by the US Department of Energy, Office of Basic Energy Sciences, contract
no. DE-SC0012704 and ARO MURI program, grant W911NF-12-1-0461.
\end{acknowledgments}

%\bibliographystyle{apsrev}
%\bibliography{PbSnTeIn}

\end{document}